\documentclass[aps,rmp,amsmath,10pt,twocolumn,superscriptaddress,showkeys,showpacs]{revtex4-1}
\usepackage[dvipdfmx]{graphicx}
\usepackage{color}
\usepackage{bm}

\begin{document}

\title{Descriptors for Machine Learning of Materials Data}
\author{Atsuto \surname{Seko}}
\email{seko@cms.mtl.kyoto-u.ac.jp}
\affiliation{Department of Materials Science and Engineering, Kyoto University}
\author{Atsushi \surname{Togo}}
\affiliation{Center for Elements Strategy Initiative for Structure Materials (ESISM), Kyoto University}
\author{Isao \surname{Tanaka}}
\affiliation{Department of Materials Science and Engineering, Kyoto University}

\date{\today}
\keywords{Machine learning interatomic potential, Lattice thermal conductivity, Recommender system, Gaussian process, Bayesian optimization}

\begin{abstract}
Descriptors, which are representations of compounds, play an essential role in machine learning of materials data.
Although many representations of elements and structures of compounds are known, these representations are difficult to use as descriptors in their unchanged forms.
This chapter shows how compounds in a dataset can be represented as descriptors and applied to machine-learning models for materials datasets.
\end{abstract}

\maketitle
\tableofcontents

\section{Introduction}
\label{nanoinfo:SecIntro}
Recent developments of data-centric approaches should accelerate the progress in materials science dramatically.
Thanks to the recent advances in computational power and techniques, the results from numerous density functional theory (DFT) calculations with predictive performances have been stored as databases.
A combination of such databases and an efficient machine-learning approach should realize prediction and classification models of target physical properties.
Consequently, machine-learning techniques are becoming ubiquitous.
They are used to explore materials and structures from a huge number of candidates and to extract meaningful information and patterns from existing data.

A key factor in controlling the performance of a machine-learning approach is how compounds are represented in a data set.
Representations of compounds are called ``descriptors'' or ``features''.
To perform machine-learning modeling, available descriptors must be determined according to the evaluation cost of the target property and the extent of the exploration space.
Based on these considerations, we aim to select ``good'' descriptors.
Prior or experts' knowledge, including a well-known correlation between the target property and the other properties, can be used to select good descriptors.
However, the set of descriptors in many cases is examined by trial-and-error because the predictive performance (i.e., the prediction error and efficiency of the model) strongly depends on the quality and data-size of the target property.

Section \ref{nanoinfo:SecCompDes} shows how to prepare descriptors of compounds.
Sections \ref{nanoinfo:SecEleRep} and \ref{nanoinfo:SecStRep} introduce representations of chemical elements (elemental representations) and atomic arrangements (structural representations) required to generate compound descriptors.
Sections \ref{nanoinfo:SecCoh}, \ref{nanoinfo:SecMLIP}, \ref{nanoinfo:SecLTC} and \ref{nanoinfo:SecRecommend} provide applications of machine-learning models for materials datasets, including the construction of a machine-learning prediction model for the DFT cohesive energy, the construction of the machine-learning interatomic potential (MLIP) for elemental metals, materials discovery of low lattice thermal conductivity (LTC), and materials discovery based on the recommender system approach.

\section{Compound descriptors}
\label{nanoinfo:SecCompDes}

Most candidate descriptors can be classified into three groups.
The first is the physical properties of a compound in a library and/or their derivative quantities, which are less available.
The second is the physical properties of a compound computed by DFT calculations or their derivative quantities.
The third is the properties of elements and the structure of a compound and/or their derivative quantities.
Combinations of different groups of descriptors can also be useful.

A set of compound descriptors should satisfy the following conditions: (i) the same-dimensional descriptors express compounds with a wide range of chemical compositions.
(ii) The same-dimensional descriptors express compounds with a wide range of crystal structures.
This is an important feature because crystals are generally composed of unit cells with different numbers of atoms.
(iii) A set of descriptors satisfies the translational, rotational, and other invariances for all compounds included in the dataset.

\begin{figure}[tbp]
\begin{center}
\includegraphics[width=\linewidth,clip]{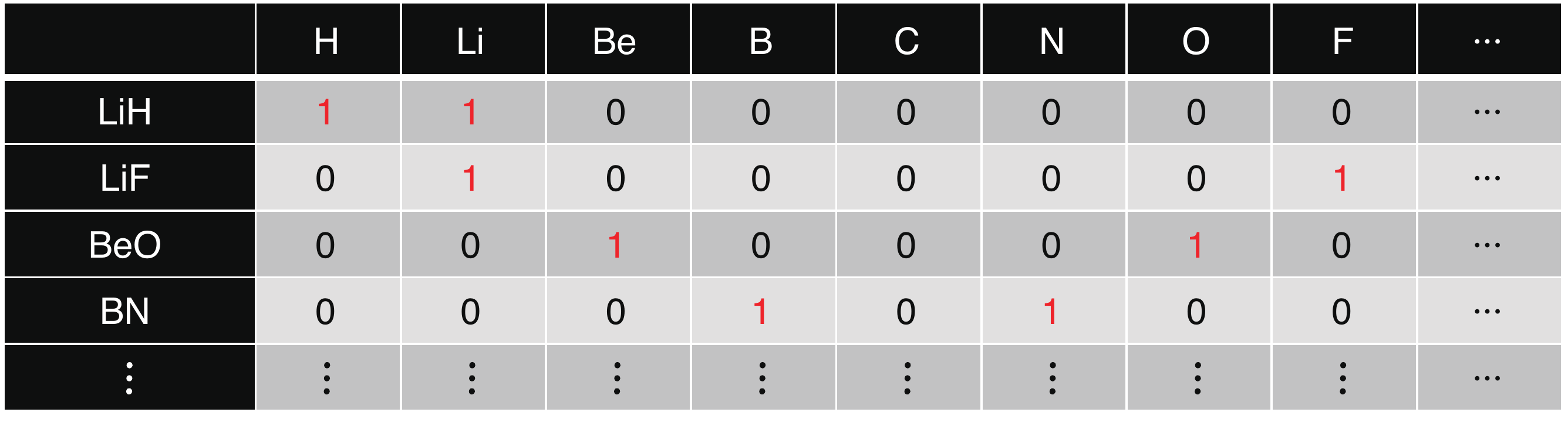} 
\caption{
Binary elemental descriptors representing the presence of chemical elements. 
The number of binary elemental descriptors corresponds to the number of element types included in the training data.
}
\label{nanoinfo:Fig-DesBinEle}
\end{center}
\end{figure}

Candidates for compound descriptors based on DFT calculations include volume, band gap, cohesive energy, elastic constants, dielectric constants, etc.
The electronic structure and phonon properties can also be used as descriptors.
Although a few first-principles databases are available, the numbers of compounds and physical properties in the databases remain limited.
Nevertheless, when a set of descriptors that can well explain a target property is discovered, a robust prediction model can be derived for the target property.
Examples can be found in the literature (e.g., Refs. \onlinecite{Fujimura_2013_AENM:AENM201300060,seko2014machine,PhysRevB.93.115104,PhysRevB.93.054112}).
Other candidates are simply a binary digit representing the presence of each element in a compound (Fig. \ref{nanoinfo:Fig-DesBinEle}) \cite{PhysRevLett.115.205901}.
When training data is composed of m kinds of elements, a compound is described by an m-dimensional binary vector with elements of one or zero.
As a simple extension, a binary digit can be replaced with the chemical composition.
Such an application is shown in Sec. \ref{nanoinfo:SecLTC} .

Another useful strategy is to use a set of quantities derived from elemental and structural representations of a compound as descriptors.
However, it is difficult to use elemental and structural representations as descriptors in their unchanged forms when the training data and search space cover a wide range of chemical compositions and crystal structures.
Consequently, it is essential to consider combined forms as compound descriptors.

\begin{figure}[tbp]
\begin{center}
\includegraphics[width=\linewidth,clip]{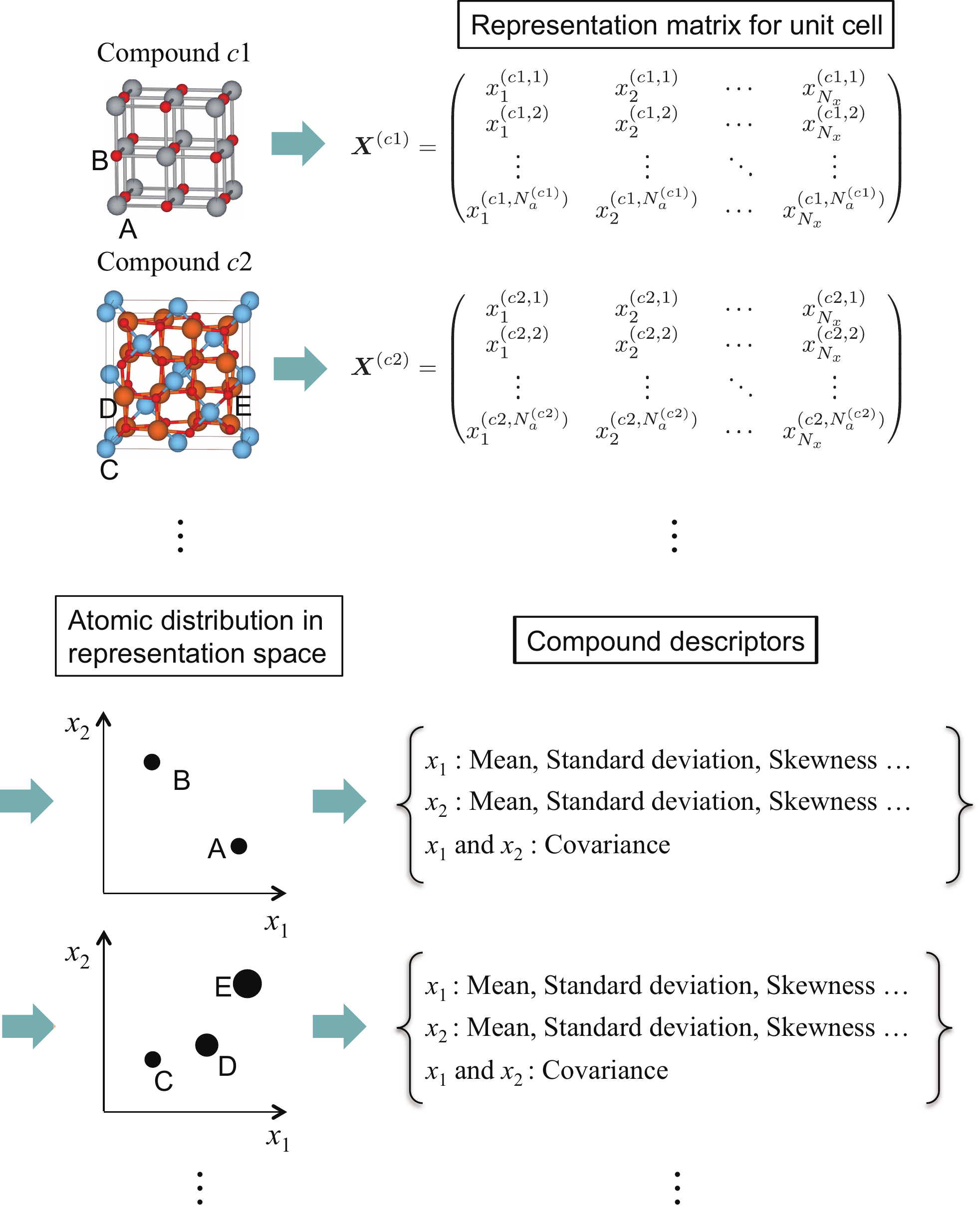} 
\caption{
Schematic illustration of how to generate compound descriptors.
}
\label{nanoinfo:Fig-Des1}
\end{center}
\end{figure}

Here we provide compound descriptors derived from elemental and structural representations satisfying the above conditions.
These descriptors can be applied not only to crystalline systems but also to molecular systems \cite{PhysRevB.95.144110}.
Figure \ref{nanoinfo:Fig-Des1} schematically illustrates the procedure to generate such descriptors for compounds.
First, the compound is considered to be a collection of atoms, which are described by element types and neighbor environments that are determined by other atoms.
Assuming the atoms are represented by $N_{x,{\rm ele}}$ elemental representations and $N_{x,{\rm st}}$ structural representations, each atom is described by $N_x = N_{x,{\rm ele}}+N_{x,{\rm st}}$ representations.
Therefore, compound $\xi$ is expressed by a collection of atomic representations as a matrix with $(N_a^{(\xi)},N_x)$-dimensions, where $N_a^{(\xi)}$ is the number of atoms in the unit cell of compound $\xi$.
The representation matrix for compound $\xi$, $\bm{X}^{(\xi)}$, is written as
\begin{equation}
\bm{X}^{(\xi)} = 
\begin{pmatrix}
x_1^{(\xi,1)} & x_2^{(\xi,1)} & \cdots & x_{N_x}^{(\xi,1)} \\
x_1^{(\xi,2)} & x_2^{(\xi,2)} & \cdots & x_{N_x}^{(\xi,2)} \\
\vdots & \vdots & \ddots & \vdots \\
x_1^{(\xi,N_a^{(\xi)})} & x_2^{(\xi, N_a^{(\xi)})} & \cdots & x_{N_x}^{(\xi,N_a^{(\xi)})} \\
\end{pmatrix}
,
\end{equation}
where $x_n^{(\xi,i)}$ denotes the $n$th representation of atom $i$ in compound $\xi$.

Since the representation matrix is only a representation for the unit cell of compound $\xi$, a procedure to transform the representation matrix into a set of descriptors is needed to compare different compounds.
One approach for this transformation is to regard the representation matrix as a distribution of data points in an $N_x$-dimensional space (Fig. \ref{nanoinfo:Fig-Des1}).
To compare the distributions themselves, representative quantities are subsequently introduced to characterize the distribution as descriptors, such as the mean, standard deviation (SD), skewness, kurtosis, and covariance.
The inclusion of the covariance enables the interaction between the element type and crystal structure to be considered.

A universal or complete set of representations is ideal because it can derive good machine-learning prediction models for all physical properties.
However, finding a universal set of representations is nearly impossible.
On the other hand, many elemental and structural representations have been proposed for a long time, not only in literature on the machine learning prediction but also in literature on the standard physics and chemistry.
Using these representations, many phenomena in physics and chemistry have been explained.
Therefore, it is a good way for generating descriptors to make effective use of the existing representations.

\section{Elemental representations}
\label{nanoinfo:SecEleRep}
The literature contains numerous quantities that can be used as elemental representations.
This chapter employs a set of elemental representations composed of the following: (1) atomic number, (2) atomic mass, (3) period and (4) group in the periodic table, (5) first ionization energy, (6) second ionization energy, (7) electron affinity, (8) Pauling electronegativity, (9) Allen electronegativity, (10) van der Waals radius, (11) covalent radius, (12) atomic radius, (13) pseudopotential radius for the s orbital, (14) pseudopotential radius for the p orbital, (15) melting point, (16) boiling point, (17) density, (18) molar volume, (19) heat of fusion, (20) heat of vaporization, (21) thermal conductivity, and (22) specific heat.
These representations can be classified into the intrinsic quantities of elements (1)-(7), the heuristic quantities of elements (8)-(14), and the physical properties of elemental substances (15)-(22).
Such elemental representations should capture essential information of compounds.
Therefore, they should assist in building models with a high predictive performance, as shown in Secs. \ref{nanoinfo:SecCoh}, \ref{nanoinfo:SecLTC}, and \ref{nanoinfo:SecRecommend}.

\section{Structural representations}
\label{nanoinfo:SecStRep}

The literature contains many structural representations that are not intended for machine learning applications.
Examples include the simple coordination number, Voronoi polyhedron of a central atom, angular distribution function, and radial distribution function (RDF).
Here, we introduce two kinds of pairwise structural representations and two kinds of angular-dependent structural representations (i.e., histogram representations of the partial radial distribution function (PRDF), generalized radial distribution function (GRDF), bond-orientational order parameter (BOP) \cite{PhysRevB.28.784}, and angular Fourier series (AFS) \cite{bartok2013representing}.

\begin{figure}[tbp]
\begin{center}
\includegraphics[width=\linewidth,clip]{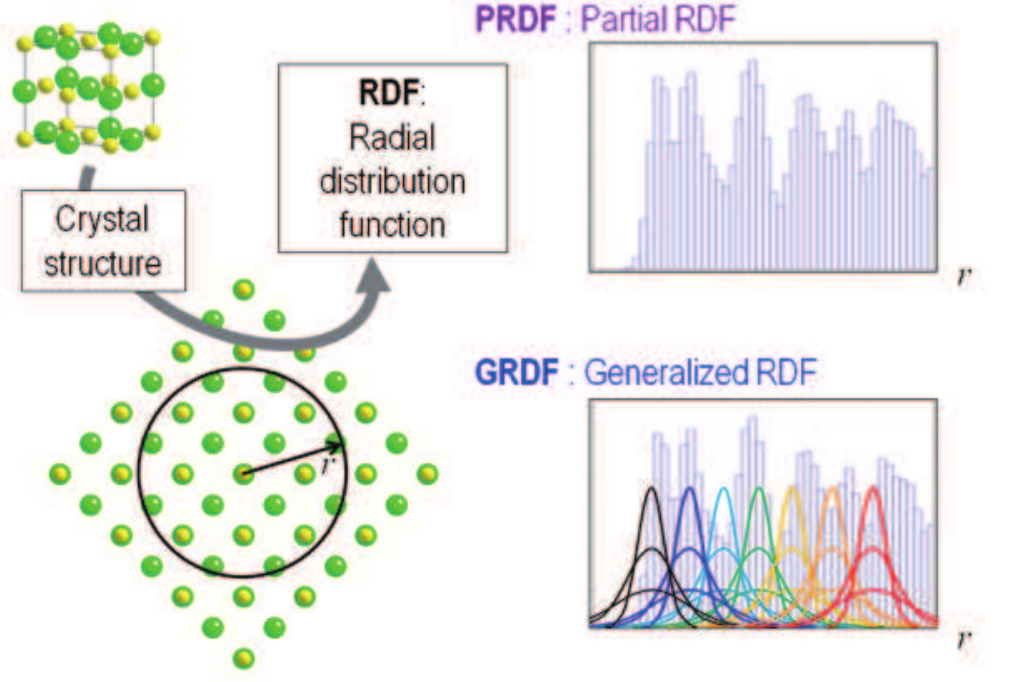} 
\caption{
Partial radial distribution functions (PRDFs) and generalized radial distribution functions (GRDFs).
}
\label{nanoinfo:Fig-DesRDF}
\end{center}
\end{figure}

The PRDF is a well-established representation for various structures.
To transform the PRDF into structural representations applicable to machine learning, a histogram representation of the PRDF is adopted with a given bin width and cutoff radius (Fig. \ref{nanoinfo:Fig-DesRDF}).
The number of counts for each bin is used as the structural representation.

The GPRF, which is a pairwise representation similar to the PRDF histogram representation, is expressed as
\begin{equation}
\label{nanoinfo:EqnGRDF}
{\rm GRDF}_n^{(i)} = \sum_j f_n (r_{ij})
\end{equation}
where $f_n (r_{ij})$ denotes a pairwise function of the distance $r_{ij}$ between atoms $i$ and $j$.
For example, a pairwise Gaussian-type function is expressed as
\begin{equation}
f_n(r) = \exp \left[-p_n (r-q_n)^2\right]f_c(r)
\end{equation}
where $f_c(r)$ denotes the cutoff function.
$p_n$ and $q_n$ are given parameters.
The GRDF can be regarded as a generalization of the PRDF histogram because the PRDF histogram is obtained using rectangular functions as pairwise functions. 

The BOP is also a well-known representation for local structures. 
The rotationally invariant BOP $Q_l^{(i)}$ for atomic neighborhoods is expressed as
\begin{equation}
Q_l^{(i)} = \left[ \frac{4\pi}{2l+1} \sum_{m=-l}^{l} |Q_{lm}^{(i)}|^2 \right]^{1/2}
\end{equation}
where $Q_{lm}^{(i)}$ corresponds to the average spherical harmonics for neighbors of atom $i$.
The third-order invariant BOP $W_l^{(i)}$ for atomic neighborhoods is expressed by 
\begin{equation}
W_l^{(i)} = \sum^{l}_{m_1, m_2, m_3 = -l}
\begin{pmatrix}
l & l & l \\
m_1 & m_2 & m_3 \\
\end{pmatrix}
Q_{lm_1}^{(i)} Q_{lm_2}^{(i)} Q_{lm_3}^{(i)},
\end{equation}
where the parentheses are the Wigner 3$j$ symbol, satisfying $m_1+m_2+m_3=0$.
A set of both $Q_l^{(i)}$ and $W_l^{(i)}$ up to a given maximum $l$ is used as the structural representations. 

The AFS is the most general among the four representations.
The AFS can include both the radial and angular dependences of an atomic distribution, and is given by 
\begin{equation}
\label{nanoinfo:EqnAFS}
{\rm AFS}_{n,l}^{(i)} = \sum_{j,k} f_n(r_{ij})f_n(r_{ik}) \cos (l \theta_{ijk})
\end{equation}
where $\theta_{ijk}$ denotes the bond angle between three atoms.

\section{Machine learning of DFT cohesive energy}
\label{nanoinfo:SecCoh}
The performances of the descriptors derived from elemental and structural representations have been examined by developing kernel ridge regression (KRR) prediction models for the DFT cohesive energy \cite{PhysRevB.95.144110}.
The dataset is composed of the cohesive energy for 18093 binary and ternary compounds computed by DFT calculations.
First, descriptor sets derived only from elemental representations, which are expected to be more dominant than structural representations in the prediction of the cohesive energy, are adopted.
Since the elemental representations are incomplete for some of the elements in the dataset, only elemental representations, which are complete for all elements, are considered.
The root-mean-square error (RMSE) is estimated for the test data.
The test data is comprised of 10\% of the randomly selected data.
This random selection of the test data is repeated 20 times, and the average RMSE is regarded as the prediction error.
 
\begin{figure}[tbp]
\begin{center}
\includegraphics[width=\linewidth,clip]{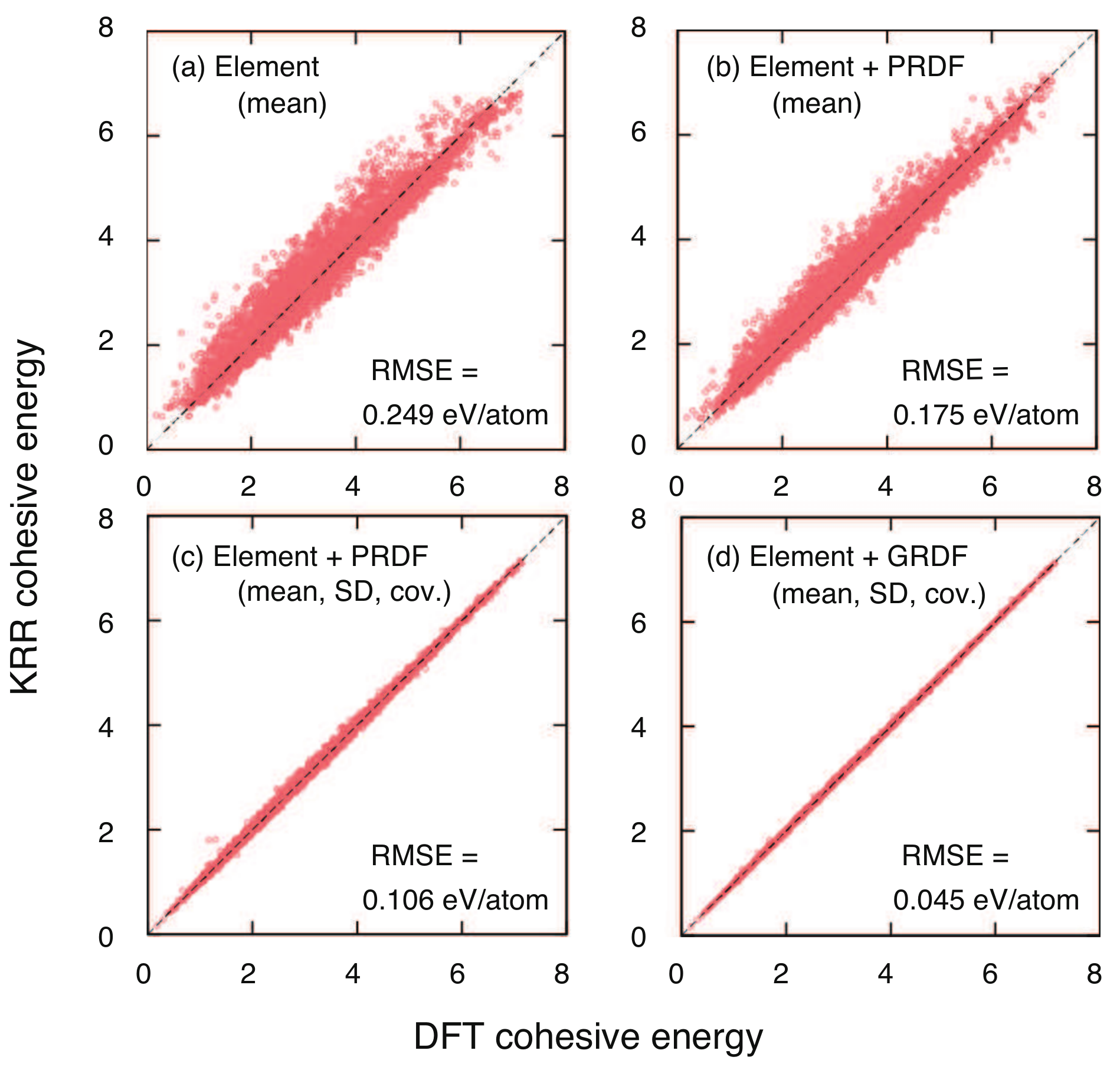} 
\caption{
Comparison of the cohesive energy calculated by DFT calculations and that calculated by the KRR prediction model.
Only one test dataset is shown.
Descriptor sets are composed of (a) the mean of the elemental representation, (b) the means of the elemental and PRDF representations, (c) the means, SDs, and covariances of the elemental and PRDF representations and (d) the means, SDs, and covariances of the elemental and 20 trigonometric GRDF representations.
Mean of the PRDF corresponds to the RDF.
Structure representations are computed from the optimized structure for each compound.
}
\label{nanoinfo:Fig-Des2}
\end{center}
\end{figure}

The simplest option is to use only the mean of each elemental representation as a descriptor.
The prediction error in this case is 0.249 eV/atom.
Figure \ref{nanoinfo:Fig-Des2} (a) compares the cohesive energy calculated by DFT calculations to that by the KRR model, where only the test data in one of the 20 trials are shown.
Numerous data points deviate from the diagonal line, which represents equal DFT and KRR energies.
When considering the means, SDs, and covariances of the elemental representations, the prediction model has a slightly smaller prediction error of 0.231 eV/atom.
Additionally, skewness and kurtosis are not important descriptors for the prediction.

Next, descriptors related to structural representations are introduced.
They can be computed from the crystal structure optimized by the DFT calculations or the initial prototype structures.
The former is only useful for machine-learning predictions when a target observation is expensive.
Since the optimized structure calculation requires the same computational cost as the cohesive energy calculation, the benefit of machine learning is lost when using the optimized structure.
The structural representations are computed from the optimized crystal structure only to examine the limitation of the procedure and representations introduced here.
KRR models are constructed using many descriptor sets, which are composed of elemental and structural representations.
The cutoff radius is set to 6 \AA\ for the PRDF, GRDF, and AFS, while the cutoff radius is set to 1.2 times the nearest-neighbor distance for the BOP.
This nearest neighbor definition is common for the BOP.

Figure \ref{nanoinfo:Fig-Des2} compares the DFT and KRR cohesive energies, where the KRR models are constructed by (b) a set of the means of the elemental and PRDF histogram representations and (c) a set of the means, standard deviations, and covariances of the elemental and PRDF histogram representations.
When considering the means of the elemental and PRDF representations, the lowest prediction error is as large as 0.166 eV/atom.
This means that simply employing the PRDF histogram does not yield a good model for the cohesive energy.
However, including the covariances of the elemental and PRDF histogram representations produces a much better prediction model and the prediction error significantly decreases to 0.106 eV/atom.

Considering only the means of the GRDFs, prediction models are obtained with errors of 0.149--0.172 eV/atom.
These errors are similar to those of prediction models considering the means of the PRDFs.
Similar to in the case of the PRDF, the prediction model improves upon considering the SDs and covariances of the elemental and structural representations.
The best model shows a prediction error of 0.045 eV/atom, which is about half that of the best PRDF model.
This is also approximately equal to the ``chemical accuracy'' of 43 meV/atom (1 kcal/mol).

Figure \ref{nanoinfo:Fig-Des2} (d) compares the DFT and KRR cohesive energies, where a set of the means, SDs, and covariances of the elemental and trigonometric GRDF representations is adopted.
Most of the data are located near the diagonal line.
We also obtain the best prediction model with a prediction error of 0.041 eV/atom by considering the means, SDs, and covariances of the elemental, 20 trigonometric GRDF, and 20 BOP representations.
Therefore, the present method should be useful to search for compounds with diverse chemical properties and applications from a wide range of chemical and structural spaces without performing exhaustive DFT calculations.

\section{Construction of MLIP for elemental metals}
\label{nanoinfo:SecMLIP}

A wide variety of conventional interatomic potentials (IPs) have been developed based on prior knowledge of chemical bonds in some systems of interest.
Examples include Lennard-Jones, embedded atom method (EAM), modified EAM (MEAM), and Tersoff potentials.
However, the accuracy and transferability of conventional IPs are often lacking due to the simplicity of their potential forms.
On the other hand, the MLIP based on a large dataset obtained by DFT calculations is beneficial to improve the accuracy and transferability.
In the MLIP framework, the atomic energy is modeled by descriptors corresponding to structural representations, as shown in Sec. \ref{nanoinfo:SecStRep}.
Once the MLIP is established, it has a similar computational cost as conventional IPs.
MLIPs have been applied to a wide range of materials, regardless of chemical bonding nature of the materials.
Recently, frameworks applicable to periodic systems have been proposed \cite{behler2007generalized,bartok2010gaussian,PhysRevB.90.024101}.

The Lasso regression has been used to derive a sparse representation for the IP.
In this section, we demonstrate the applicability of the Lasso regression to derive the IPs of 12 elemental metals (Na, Mg, Ag, Al, Au, Ca, Cu, Ga, In, K, Li, and Zn) \cite{PhysRevB.90.024101,PhysRevB.92.054113}.
The features of linear modeling of the atomic energy and descriptors using the Lasso regression include the following.
1) The accuracy and computational cost of the energy calculation can be controlled in a transparent manner.
2) A well-optimized sparse representation for the IP, which can accelerate and increase the accuracy of atomistic simulations while decreasing the computational costs, is obtained.
3) Information on the forces acting on atoms and stress tensors can be included in the training data in a straightforward manner.
4) Regression coefficients are generally determined quickly using the standard least-squares technique.

The total energy of a structure can be regarded as the sum of the constituent atomic energies.
In the framework of MLIPs with only pairwise descriptors, the atomic energy of atom $i$ is formulated as
\begin{equation}
E^{(i)} = F\left(b_1^{(i)}, b_2^{(i)}, \cdots, b_{n_{\rm max}}^{(i)} \right),
\end{equation}
where $b_n^{(i)}$ denotes a pairwise descriptor.
Numerous pairwise descriptors are generally used to formulate the MLIP.
We use the GRDF expressed by Eqn.(\ref{nanoinfo:EqnGRDF}) as the descriptors.
For the pairwise function $f_n$, we introduce Gaussian, cosine, Bessel, Neumann, modified Morlet wavelet, Slater-type orbital, and Gaussian-type orbital functions.
Although artificial neural network and Gaussian process black-box models have been used as functions $F$, we use a polynomial function to construct the MLIPs for the 12 elemental metals.
In the approximation considering only the power of $b_n^{(i)}$, the atomic energy is expressed as
\begin{equation}
E^{(i)} = w_0 + \sum_n w_n b_n^{(i)} + \sum_n w_{n,n} b_n^{(i)}b_n^{(i)} + \cdots,
\end{equation}
where $w_0$, $w_n$, and $w_{n,n}$ denote the regression coefficients. Practically, the formulation is truncated by the maximum value of power, $p_{\rm max}$.

The vector $\bm{w}$ composed of all the regression coefficients can be estimated by a regression, which is a machine learning method to estimate the relationship between the predictor and observation variables using a training dataset.
For the training data, the energy, forces acting on atoms, and stress tensor computed by DFT calculations can be used as the observations in the regression process since they all are expressed by linear equations with the same regression coefficients \cite{PhysRevB.92.054113}.
A simple procedure to estimate the regression coefficients employs a linear ridge regression \cite{hastieelements}.
This is a shrinkage method where the number of regression coefficients is reduced by imposing a penalty.
The ridge coefficients minimize the penalized residual sum of squares and are expressed as 
\begin{equation}
L(\bm{w}) = ||\bm{X}\bm{w} - \bm{y}||^2_2 + \lambda ||\bm{w}||^2_2, 
\end{equation}
where $\bm{X}$ and $\bm{y}$ denote the predictor matrix and observation vector, respectively, which correspond to the training data.
$\lambda$, which is called the regularization parameter, controls the magnitude of the penalty.
This is referred to as L2 regularization.
The regression coefficients can easily be estimated while avoiding the well-known multicollinearity problem that occurs in the ordinary least-squares method.

Although the linear ridge regression is useful to obtain an IP from a given descriptor set, a set of descriptors relevant to the system of interest is generally unknown.
Moreover, an MLIP with a small number of descriptors is desirable to decrease the computational cost in atomistic simulations.
Therefore, a combination of the Lasso regression \cite{hastieelements,tibshirani1996regression} and a preparation involving a considerable number of descriptors is used.
The Lasso regression provides a solution to the linear regression as well as a sparse representation with a small number of non-zero regression coefficients.
The solution is obtained by minimizing the function that includes the L1 norm of regression coefficients and is expressed as 
\begin{equation}
L(\bm{w}) = ||\bm{X}\bm{w} - \bm{y}||^2_2 + \lambda ||\bm{w}||_1.
\end{equation}
Simply adjusting the values of $\lambda$ for a given training dataset controls the accuracy of the solution.

To begin with, training and test datasets are generated from DFT calculations.
The test dataset is used to examine the predictive power for structures that are not included in the training dataset.
For each elemental metal, 2700 and 300 configurations are generated for the training and test datasets, respectively.
The datasets include structures made by isotropic expansions, random expansions, random distortions, and random displacements of ideal face-centered-cubic (fcc), body-centered-cubic (bcc), hexagonal-closed-packed (hcp), simple-cubic (sc), $\omega$ and $\beta$-tin structures, in which the atomic positions and lattice constants are fully optimized.
These configurations are made using supercells constructed by the $2\times2\times2$, $3\times3\times3$, $3\times3\times3$, $4\times4\times4$, $3\times3\times3$ and $2\times2\times2$ expansions of the conventional unit cells for fcc, bcc, hcp, sc, $\omega$, and $\beta$-tin structures, which are composed of 32, 54, 54, 64, 81 and 32 atoms, respectively.

For a total of 3000 configurations for each elemental metal, DFT calculations have been performed using the plane-wave basis projector augmented wave (PAW) method \cite{PAW1} within the Perdew--Burke--Ernzerhof exchange-correlation functional \cite{GGA:PBE96} as implemented in the VASP code \cite{VASP1,VASP2,PAW2}.
The cutoff energy is set to 400 eV.
The total energies converge to less than 10$^{-3}$ meV/supercell.
The atomic positions and lattice constants are optimized for the ideal structures until the residual forces are less than 10$^{-3}$ eV/\AA.

\begin{table}[tbp]
\caption{
RMSEs for the test data of linear ridge MLIPs using 240 terms. (Unit: meV/atom)
}
\label{nanoinfo:Table-IP1}
\begin{ruledtabular}
\begin{tabular}{lcc}
Function type for $f_n$ and $p_{\rm max}$ & Na & Mg \\
\hline
Cosine $(p_{\rm max} = 1)$ & 7.3 & 11.8   \\
Cosine $(p_{\rm max} = 2)$ & 1.6 & 2.6   \\
Cosine $(p_{\rm max} = 3)$ & 1.4 & 1.6   \\
Cosine, Gaussian $(p_{\rm max} = 3)$ & 1.4 & 1.1  \\
Cosine, Bessel $(p_{\rm max} = 3)$ & 1.4 & 1.3  \\
Cosine, Gaussian, Bessel $(p_{\rm max} = 3)$ & 1.4 & 0.9  \\
\end{tabular}
\end{ruledtabular}
\end{table}

For each MLIP, the RMSE is calculated between the energies for the test data predicted by the DFT calculations and those predicted using the MLIP.
This can be regarded as the prediction error of the MLIP.
Table \ref{nanoinfo:Table-IP1} shows the RMSEs of linear ridge MLIPs with 240 terms for Na and Mg, where the RMSE converges as the number of terms increases.
The MLIPs with only pairwise interactions have low predictive powers for both Na and Mg.
Increasing pmax improves the predictive power of the MLIPs substantially.
Using cosine-type functions with $p_{\rm max} = 3$ and cutoff radius $R_c = 7.0$ \AA, the RMSEs are 1.4 and 1.6 meV/atom for Na and Mg, respectively.
By increasing the cutoff radius to $R_c = 9.0$ \AA, the RMSE reaches a very small value of 0.4 meV/atom for Na, but the RMSE remains almost unchanged for Mg.
The RMSE for Na is not improved, even after considering all combinations of the Gaussian, cosine, Bessel and Neumann descriptor sets.
In contrast, the combination of Gaussian, cosine, and Bessel descriptor sets provides the best prediction for Mg with an RMSE of 0.9 meV/atom.

\begin{figure}[tbp]
\begin{center}
\includegraphics[width=\linewidth,clip]{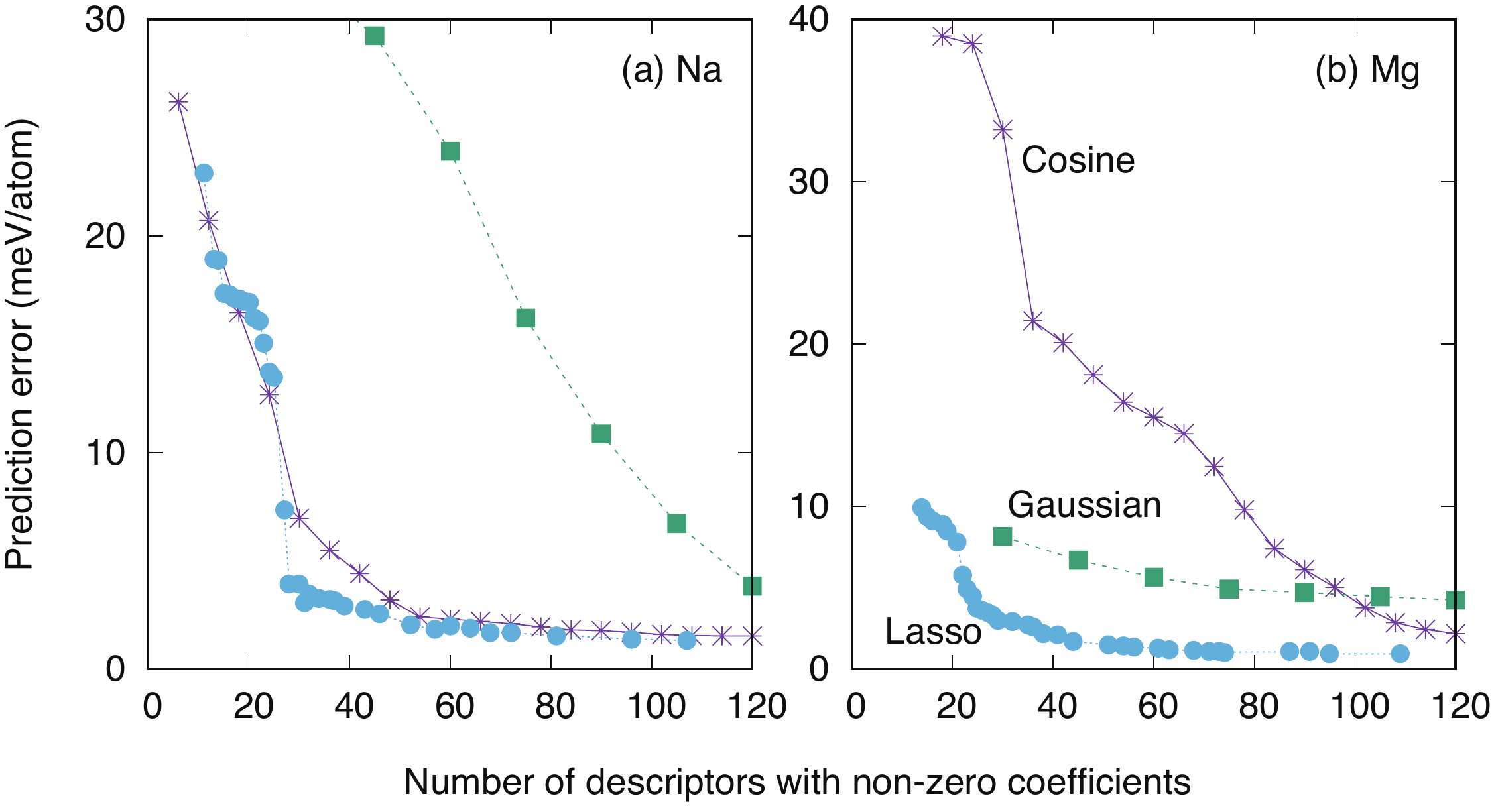} 
\caption{
RMSEs for the test data of the linear ridge MLIP using cosine-type and Gaussian-type descriptors with $p_{\rm max} = 3$, $R_c = 7.0$ \AA\ and $\lambda=0.001$ for (a) Na and (b) Mg. 
RMSEs of the Lasso MLIPs are also shown. 
}
\label{nanoinfo:Fig-LassoIP1}
\end{center}
\end{figure}

The Lasso MLIPs have been constructed using the same dataset.
Candidate terms for the Lasso MLIPs are composed of numerous Gaussian, cosine, Bessel, Neumann, polynomial and GTO descriptors.
Sparse representations are then extracted from a set of candidate terms by the Lasso regression.
Figure \ref{nanoinfo:Fig-LassoIP1} shows the RMSEs of the Lasso MLIPs for Na and Mg, respectively.
The RMSEs of the Lasso MLIP decrease faster than those of the linear ridge MLIPs constructed from a single-type of descriptors.
In other words, the Lasso MLIP requires fewer terms than the linear ridge MLIP.
For Na, a sparse representation with an RMSE of 1.3 meV/atom is obtained using only 107 terms.
This is almost the same accuracy as the linear ridge MLIP with 240 terms based on the cosine descriptors.
It is apparent that the Lasso MLIP is more advantageous for Mg than for Na.
The obtained sparse representation with 95 terms for Mg has an RMSE of 0.9 meV/atom.
This is almost half the terms for the linear ridge MLIP based on the cosine descriptors, which requires 240 terms.

\begin{figure*}[tbp]
\begin{center}
\includegraphics[width=0.8\linewidth,clip]{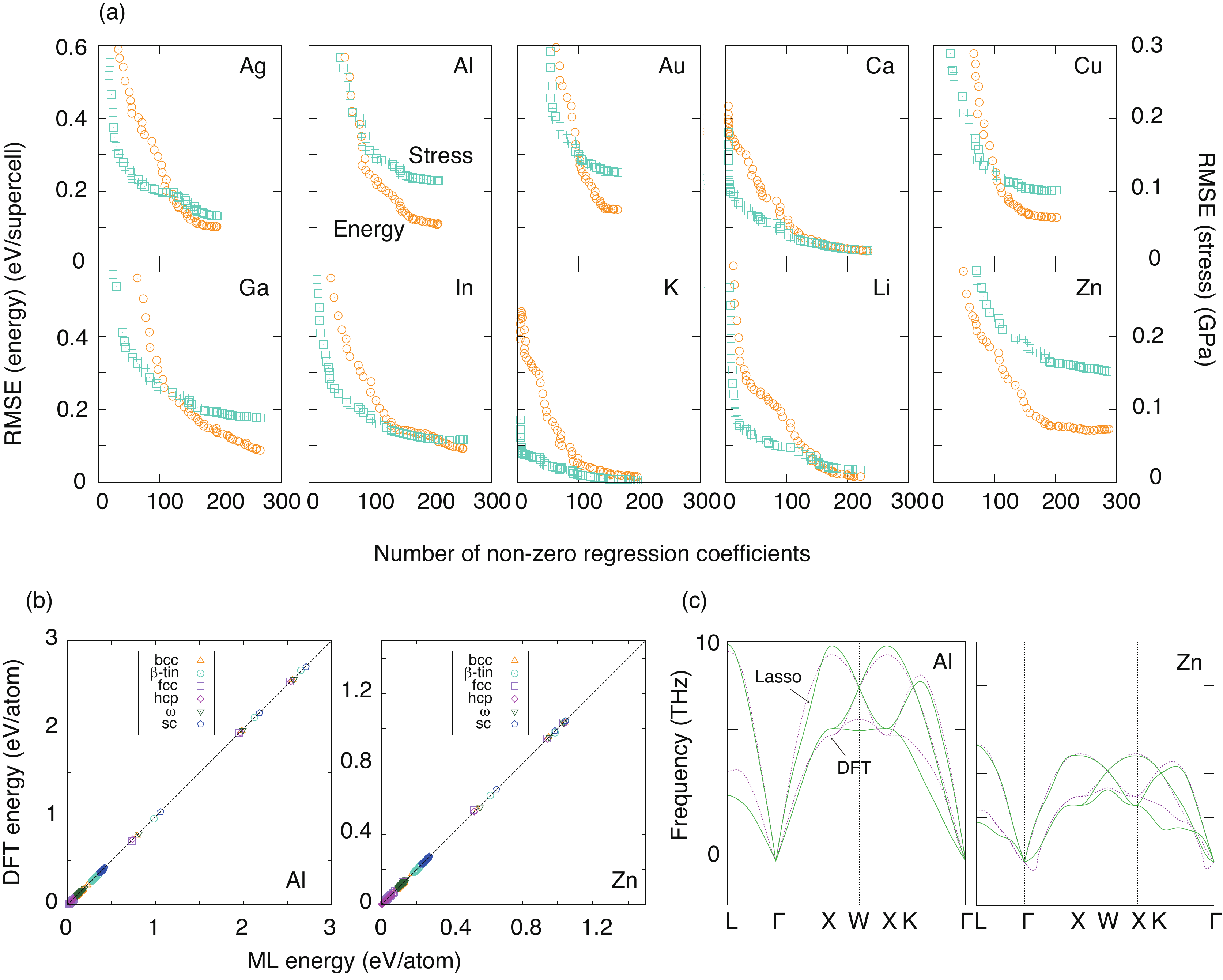} 
\caption{
(a) Dependence of RMSEs for the energy and stress tensor of the Lasso MLIP on the number of non-zero regression coefficients for ten elemental metals. Orange open circles and blue open squares show RMSEs for the energy and stress tensor, respectively. (b) Comparison of the energies predicted by the Lasso MLIP and DFT for Al and Zn measured from the energy of the most stable structure. (c) Phonon dispersion relationships for FCC-Al and FCC-Zn. Blue solid and orange broken lines show the phonon dispersion curves obtained by the Lasso MLIP and DFT, respectively. Negative values indicate imaginary modes. 
}
\label{nanoinfo:Fig-LassoIP2}
\end{center}
\end{figure*}

\begin{table*}[tbp]
\caption{
RMSEs for the energy, force, and stress tensor of the Lasso MLIPs showing the minimum criterion score. 
Optimal cutoff radius for each element is also shown.
}
\label{nanoinfo:Table-IP2}
\begin{ruledtabular}
\begin{tabular}{cccccc}
Element & Cutoff radius & Number of  & RMSE (energy) & RMSE (force) & RMSE (stress) \\
& (\AA) & basis functions & (meV/atom) & (eV/\AA) & (GPa)  \\
\hline
Ag & 7.5 & 190 & 2.2 & 0.011 & 0.07  \\
Al & 8.0  & 210 & 3.5 & 0.020 & 0.12  \\
Au & 6.0  & 165 & 2.4 & 0.030 & 0.15  \\
Ca & 9.5  & 234 & 1.2 & 0.010 & 0.03  \\
Cu & 7.5  & 202 & 2.6 & 0.018 & 0.12  \\
Ga & 10.0 & 266 & 2.2 & 0.017 & 0.09  \\
In & 10.0 & 253 & 2.3 & 0.019 & 0.07  \\
K  & 10.0 & 197 & 0.3 & 0.001 & 0.00  \\
Li & 8.5  & 222 & 0.4 & 0.005 & 0.02  \\
Zn & 10.0 & 288 & 2.9 & 0.016 & 0.15  \\
\end{tabular}
\end{ruledtabular}
\end{table*}

Figure \ref{nanoinfo:Fig-LassoIP2} (a) shows the dependence of the RMSE for the energy and stress tensor of the Lasso MLIP on the number of non-zero regression coefficients for the other ten elemental metals.
The number of selected terms tends to increase as the regularization parameter $\lambda$ decreases.
The RMSEs for the energy and stress tensor tend to decrease.
Although multiple MLIPs with the same number of terms are sometimes obtained from different values of $\lambda$, only the MLIP with the lowest criterion score with the same number of terms is shown in Fig. \ref{nanoinfo:Fig-LassoIP2} (a).
Table \ref{nanoinfo:Table-IP2} shows the RMSEs for the energy, force, and stress tensor of the optimal Lasso MLIP.
The MLIPs are obtained with the RMSE for the energy in the range of 0.3--3.5 meV/atom for the ten elemental metals using only 165--288 terms.
The RMSEs for the force and stress are within 0.03 eV/\AA\ and 0.15 GPa, respectively.

Figure \ref{nanoinfo:Fig-LassoIP2} (b) compares the energies of the test data predicted by the Lasso MLIP and DFT for Al and Zn.
Both the largest and second largest RMSEs for the energy are shown.
Regardless of the crystal structure, the DFT and Lasso MLIP energies are similar.
In addition, the RMSE is clearly independent of the energy despite the wide range of structures included in both the training and test data.

The applicability of the Lasso MLIP to the calculation of the force has been also examined by comparing the phonon dispersion relationships computed by the Lasso MLIP and DFT.
The phonon dispersion relationships are calculated by the supercell approach for the fcc structure with the equilibrium lattice constant.
The phonon calculations use the phonopy code \cite{togo2015first}.
Figure \ref{nanoinfo:Fig-LassoIP2} (c) shows the phonon dispersion relationships of the fcc structure for elemental Al and Zn computed by both the Lasso MLIP and DFT.
The phonon dispersion relationships calculated by the Lasso MLIP agree well with those calculated by DFT.
This demonstrates that the Lasso MLIP is sufficiently accurate to perform atomistic simulations with an accuracy similar to DFT calculations.

It is important to use an extended approximation for the atomic energy in transition metals \cite{takahashi2017conceptual}.
The extended approximation also improves the predictive power for the above elemental metals.
The MLIPs are constructed by a second-order polynomial approximation with the AFSs described by Eqn.(\ref{nanoinfo:EqnAFS}) and their cross terms.
For elemental Ti, the optimized angular-dependent MLIP is obtained with a prediction error of 0.5 meV/atom (35245 terms), which is much smaller than that of the Lasso MLIP with only the power of pairwise descriptors of 17.0 meV/atom.
This finding demonstrates that it is very important to consider angular-dependent descriptors when expressing interatomic interactions of elemental Ti.
The angular-dependent MLIP can predict the physical properties much more accurately than existing IPs.

\section{Discovery of low lattice thermal conductivity materials}
\label{nanoinfo:SecLTC}

Thermoelectric generators are essential to utilize waste heat.
The thermoelectric figure of merit should be increased to improve the conversion efficiency.
Since the figure of merit is inversely proportional to the thermal conductivity, many works have strived to reduce the thermal conductivity, especially the LTC.
To evaluate LTCs with an accuracy comparable to the experimental data, a method that greatly exceeds ordinary DFT calculations is required.
Since multiple interactions among phonons, or anharmonic lattice dynamics, must be treated, the computational cost is many orders of magnitudes higher than ordinary DFT calculations of primitive cells.
Such expensive calculations are feasible only for a few simple compounds.
High-throughput screening of a large DFT database of the LTC is an unrealistic approach unless the exploration space is narrowly confined.

\begin{figure}[tbp]
\begin{center}
\includegraphics[width=\linewidth,clip]{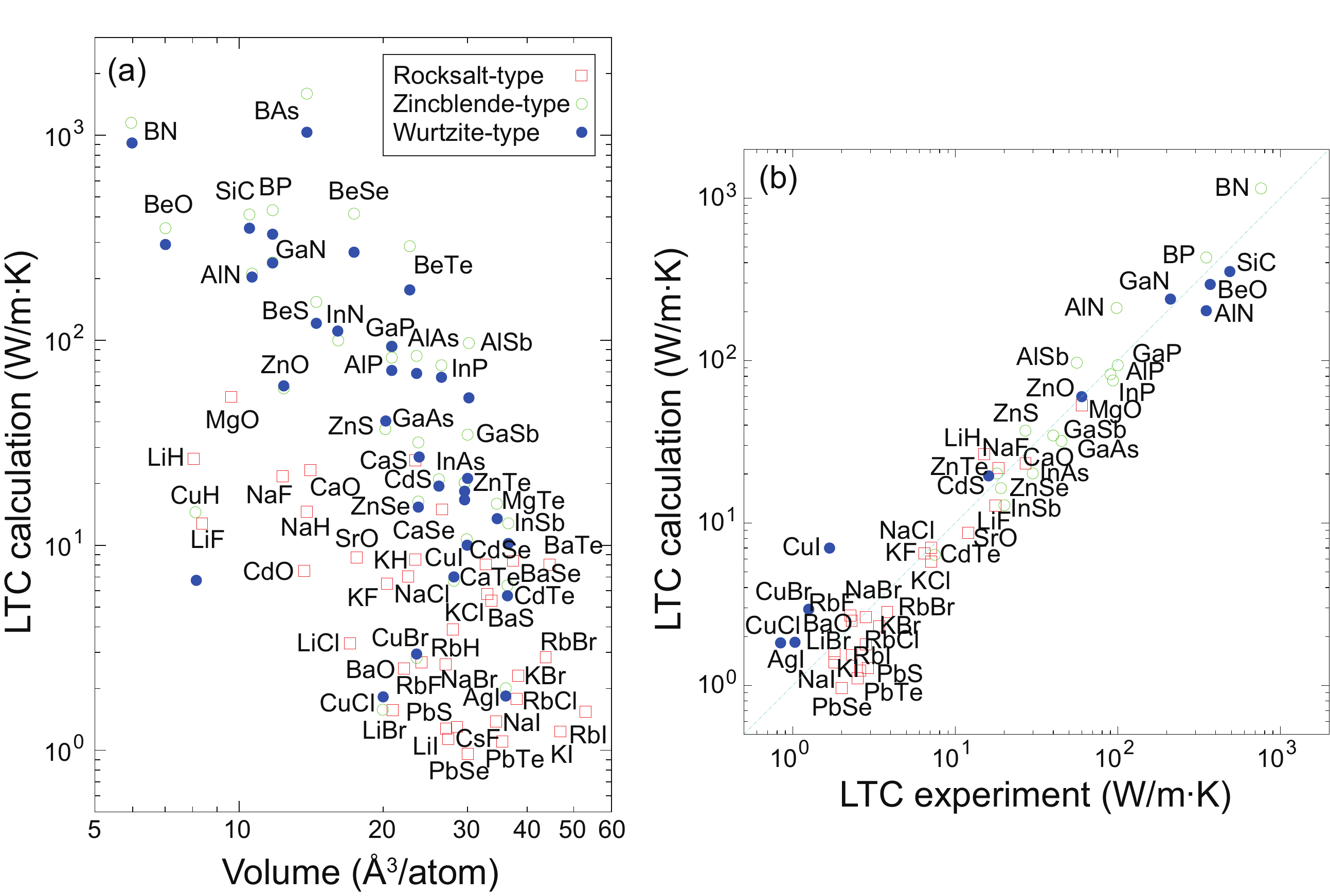} 
\caption{
(a) LTC calculated from the first principles calculations for 101 compounds along with volume, $V$. 
(b) Experimental LTC data are shown for comparison when the experimental LTCs are available.
}
\label{nanoinfo:Fig-LTC1}
\end{center}
\end{figure}

Recently, Togo et al. reported a method to systematically obtain the theoretical LTC through first principles anharmonic lattice dynamics calculations \cite{togo2015distributions}.
Figure \ref{nanoinfo:Fig-LTC1} (a) shows the results of first-principles LTCs for 101 compounds as functions of the crystalline volume per atom, $V$.
PbSe with the rocksalt structure shows the lowest LTC, 0.9 W/mK (at 300 K).
Its trend is similar to that in a recent report on low LTC for lead- and tin-chalcogenides.

Figure \ref{nanoinfo:Fig-LTC1} (b) compares the computed results with the available experimental data.
The satisfactory agreement between the experimental and computed results demonstrates the usefulness of the first-principles LTC data for further studies.
A phenomenological relationship has been proposed where $\log \kappa_L$ is proportional to $\log V$ \cite{slack1979thermal}.
Although a qualitative correlation is observed between our LTC and $V$, it is difficult to predict the LTC quantitatively or discover new compounds with low LTCs only from the phenomenological relationship.
It should be noted that the dependence on $V$ differs remarkably between rocksalt-type and zincblende- or wurtzite-type compounds.
However, zincblende- and wurtzite-type compounds show a similar LTC for the same chemical composition.
The 101 first-principles LTC data has been used to create a model to predict the LTCs of compounds within a library \cite{PhysRevLett.115.205901}.
Firstly, a Gaussian process (GP)-based Bayesian optimization \cite{Rasmussen_2006} is adopted using two physical quantities as descriptors: $V$ and density, $\rho$.
These quantities are available in most experimental or computational crystal structure databases.
Although a phenomenological relationship is proposed between $\log \kappa_L$ and $V$, the correlation between them is low.
Moreover, the correlation between $\log \kappa_L$ and $\rho$ is even worse.

We start from an observed data set of five compounds that are randomly chosen from the dataset.
The Bayesian optimization searches for the compound with a maximum probability of improvement \cite{jones01} among the remaining data.
That is, the compound with the highest Z-score derived from GP is searched.
The compound is included into the observed dataset.
Then another compound with the maximum probability of improvement is searched.
Both the Bayesian optimization and random searches are repeated 200 times, and the average number of observed compounds required to find the best compound is examined.

The average numbers of compounds required for the optimization using the Bayesian optimization and random searches, $N_{\rm ave}$, are 11 and 55, respectively.
The compound with the lowest LTC among the 101 compounds (i.e., rocksalt PbSe) can be found much more efficiently using a Bayesian optimization with only two variables, $V$ and $\rho$.
However, using a Bayesian optimization only with these two variables is not a robust method to determine the lowest LTC.
As an example, the result of the Bayesian optimization using the dataset after intentionally removing the first and second lowest LTC compounds shows that $N_{\rm ave}$ is 65 to find LiI using Bayesian optimization only with $V$ and $\rho$, which is larger than that of the random search ($N_{\rm ave} = 50$).
The delay in the optimization should originate from the fact that LiI is an outlier when the LTC is modeled only with $V$ and $\rho$.
Such outlier compounds with low LTC are difficult to find only with $V$ and $\rho$.

To overcome the outlier problem, predictors have been added for constituent chemical elements.
There are many choices for such variables.
Here, we introduce binary elemental descriptors, which are a set of binary digits representing the presence of chemical elements.
Since the 101 LTC data is composed of 34 kinds of elements, there are 34 elemental descriptors.
When finding both PbSe and LiI, the compound with the lowest LTC is found with $N_{\rm ave} = 19$.
The use of binary elemental descriptors improves the robustness of the efficient search.

\begin{figure}[tbp]
\begin{center}
\includegraphics[width=\linewidth,clip]{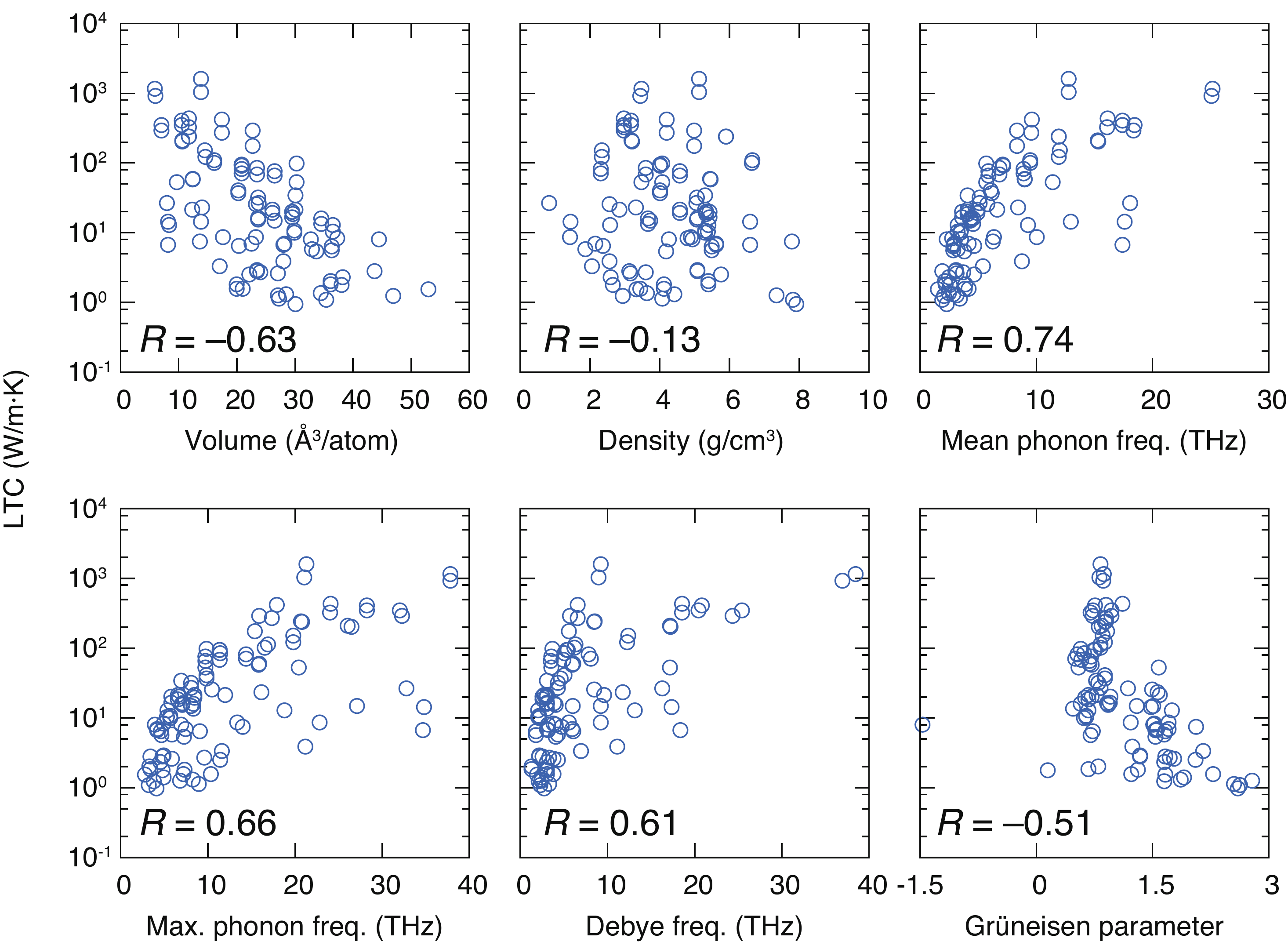} 
\caption{
Relationship between $\log \kappa_L$ and the physical properties derived from the first principles electronic structure and phonon calculations.
Correlation coefficient, $R$, is shown in each panel. 
}
\label{nanoinfo:Fig-LTC2}
\end{center}
\end{figure}

Better correlations with LTC can be found for parameters obtained from the phonon density of states.
Figure \ref{nanoinfo:Fig-LTC2} shows the relationships between the LTC and the physical properties.
Other than volume and density, the following quantities are obtained by our phonon calculations: mean phonon frequency, maximum phonon frequency, Debye frequency, and Gr\"uneisen parameter.
The Debye frequency is determined by fitting the phonon density of states for a range between 0 and 1/4 of the maximum phonon frequency to a quadratic function.
The thermodynamic Gr\"uneisen parameter is obtained from the mode-Gr\"uneisen parameters calculated with a quasi-harmonic approximation and mode-heat capacities.
The correlation coefficients $R$ between $\log \kappa_L$ and these physical properties are shown in the corresponding panels.
The present study does not use such phonon parameters as descriptors because a data library for such phonon parameters for a wide range of compounds is unavailable.
Hereafter, we show results only with the descriptor set composed of 34 binary elemental descriptors on top of $V$ and $\rho$.

A GP prediction model has been used to screen for low LTC compounds in a large library of compounds.
In the biomedical community, a screening based on a prediction model is called a ``virtual screening'' \cite{kitchen2004docking}.
For the virtual screening, all 54779 compounds in the Materials Project Database (MPD) library \cite{jain2013commentary}, which is composed mostly of crystal structure data available in ICSD \cite{bergerhoff1987crystal}, are adopted.
Most of these compounds have been synthesized experimentally at least once.
On the basis of the GP prediction model made by $V$, $\rho$, and the 34 binary elemental descriptors for the 101 LTC data, low-LTC compounds are ranked according to the Z-score of the 54779 compounds.

\begin{figure*}[tbp]
\begin{center}
\includegraphics[width=0.8\linewidth,clip]{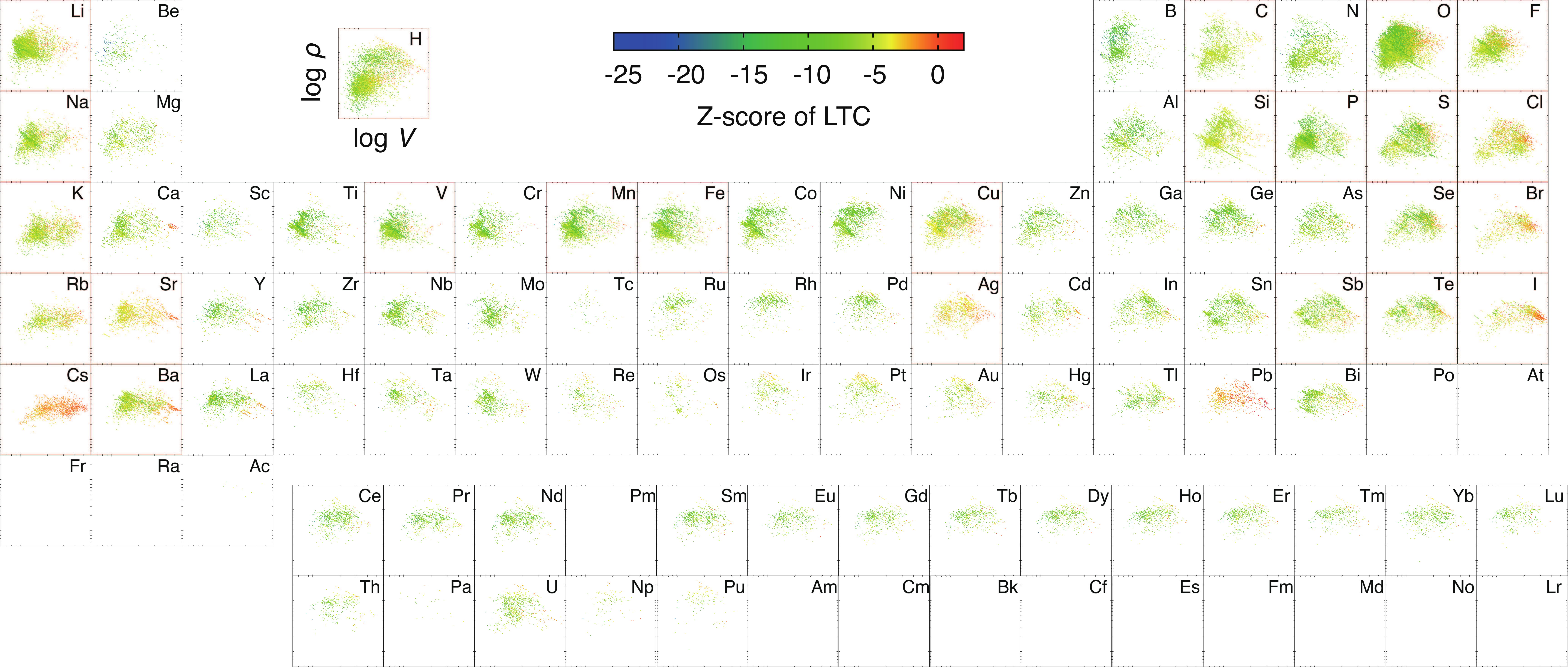} 
\caption{
Dependence of the Z-score on the constituent elements for compounds in the MPD library.
Color along the volume and density for each element denote the magnitude of the Z-score. 
}
\label{nanoinfo:Fig-LTC3}
\end{center}
\end{figure*}

Figure \ref{nanoinfo:Fig-LTC3} shows the distribution of Z-scores for the 54779 compounds along with $V$ and $\rho$.
The magnitude of the Z-score is plotted in the panels corresponding to the constituent elements.
The compounds are widely distributed in $V-\rho$ space.
Thus, it is difficult to identify compounds without performing a Bayesian optimization with elemental descriptors.
The widely distributed Z-scores for light elements such as Li, N, O, and F imply that the presence of such light elements has a negligible effect on lowering the LTC.
When such light elements form a compound with heavy elements, the compound tends to show a high Z-score.
It is also noteworthy that many compounds composed of light elements such as Be and B tend to show a high LTC.
Pb, Cs, I, Br, and Cl exhibit special features.
Many compounds composed of these elements exhibit high Z-scores.
Most compounds showing a positive Z-score are a combination of these five elements.
On the other hand, elements in the periodic table neighboring these five elements do not show analogous trends.
For example, compounds of Tl and Bi, which neighbor Pb, rarely exhibit high Z-scores.
This may sound odd since Bi$_2$Te$_3$ is a famous thermoelectric compound, and some compounds containing Tl have a low LTC.
This may be ascribed to our selection of the training dataset, which is composed only of AB compounds with 34 elements and three kinds of simple crystal structures.
In other words, the training dataset is somehow ``biased''. 
Currently, this bias is unavoidable because first-principles LTC calculations are still too expensive to obtain a sufficiently unbiased training dataset with a large enough number of data points to cover the diversity of the chemical compositions and crystal structures.
Nevertheless, the usefulness of biased training dataset to find low-LTC materials will be verified in the future.
Due to the biased training dataset, all low-LTC materials in the library may not be discovered.
However, some of them can be discovered.
A ranking of LTCs from the Z-score does not necessarily correspond to the true first-principles ranking.
Therefore, a verification process for candidates of low-LTC compounds after the virtual screening is one of the most important steps in ``discovering'' low-LTC compounds.
First principles LTCs have been evaluated for the top eight compounds after the virtual screening.
All of them are considered to form ordered structures.
However, the LTC calculation is unsuccessful for Pb$_2$RbBr$_5$ due to the presence of imaginary phonon modes within the supercell used in the present study.
All of the top five compounds, PbRbI$_3$, PbIBr, PbRb$_4$Br$_6$, PbICl and PbClBr, show a LTC of $<$ 0.2 W/mK (at 300 K), which is much lower than that of the rocksalt PbSe, [i.e., 0.9 W/mK (at 300 K)].
This confirms the powerfulness of the present GP prediction model to efficiently discover low-LTC compounds.
The present method should be useful to search for materials in diverse applications where the chemistry of materials must be optimized.

\begin{figure}[tbp]
\begin{center}
\includegraphics[width=\linewidth,clip]{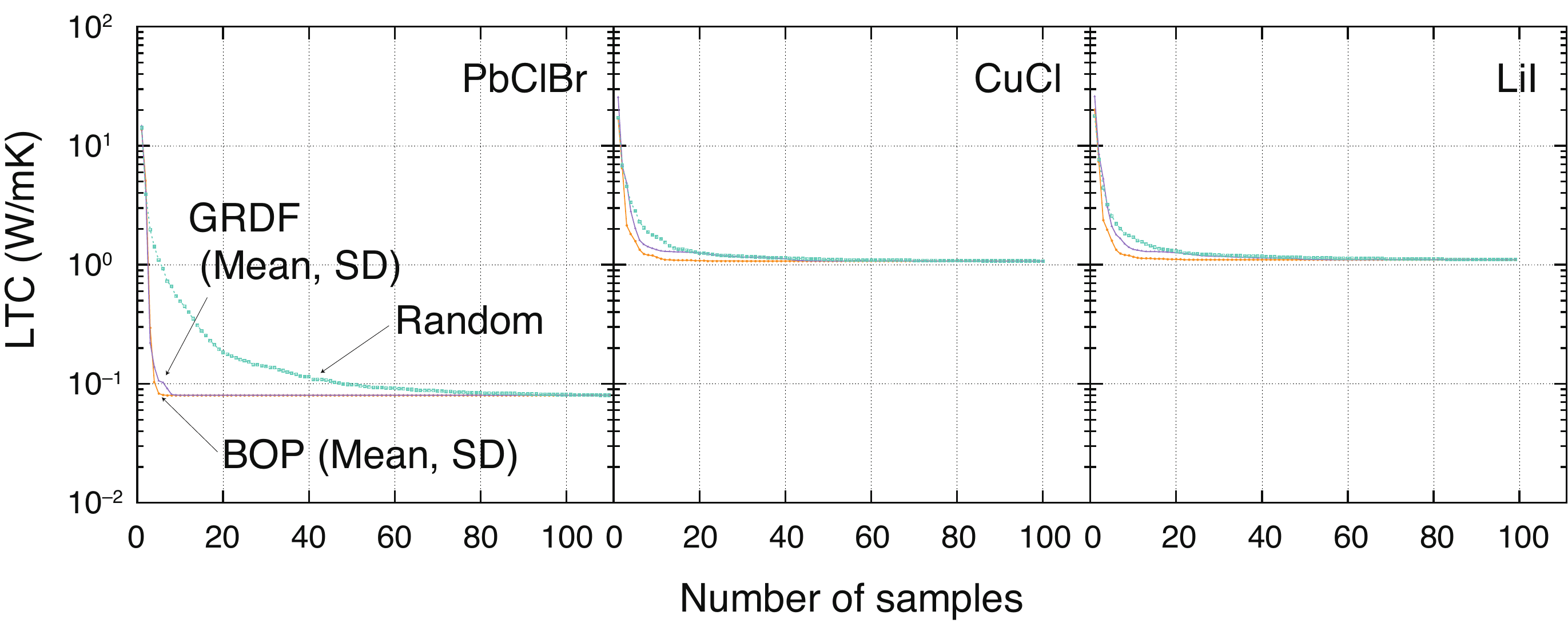} 
\caption{
Behavior of the Bayesian optimization for the LTC data to find PbClBr, CuCl, and LiI. 
}
\label{nanoinfo:Fig-LTC4}
\end{center}
\end{figure}

Finally, the performance of Bayesian optimization has been examined using the compound descriptors derived from elemental and structural representations for the LTC dataset containing the compounds identified by the virtual screening.
GP models are constructed using (1) the means and SDs of the elemental representations and GRDFs and (2) the means and SDs of elemental representations and BOPs.
Figure \ref{nanoinfo:Fig-LTC4} shows the behavior of the lowest LTC during Bayesian optimization relative to a random search.
The optimization aims to find PbClBr with the lowest LTC.
For the GP model with the BOP, the average number of samples required for the optimization, $N_{\rm ave}$, is 5.0, which is ten times smaller than that of the random search, $N_{\rm ave} = 50$.
Hence, the Bayesian optimization more efficiently discovers PbClBr than the random search.

To evaluate the ability to find a wide variety of low-LTC compounds, two datasets have been prepared after intentionally removing some low-LTC compounds.
In these datasets, CuCl and LiI, which respectively show the 11th-lowest and 12th-lowest LTCs, are solutions of the optimizations.
For the GP model with BOPs, the average number of observations required to find CuCl and LiI is  $N_{\rm ave} = 15.1$ and 9.1, respectively.
These numbers are much smaller than those of the random search.
On the other hand, for the GP model with GRDFs, the average number of observations required to find CuCl and LiI is $N_{\rm ave} = 40.5$ and 48.6, respectively.
The delayed optimization may originate from the fact that both CuCl and LiI are outliers in the model with GRDFs, although the model with GRDFs has a similar RMSE as the model with BOPs.
These results indicate that the set of descriptors needs to be optimized by examining the performance of Bayesian optimization for a wide range of compounds to find outlier compounds.

\section{Recommender system approach for materials discovery}
\label{nanoinfo:SecRecommend}
Many atomic structures of inorganic crystals have been collected.
Of the few available databases for inorganic crystal structures, the ICSD \cite{bergerhoff1987crystal} contains approximately 10$^5$ inorganic crystals, excluding duplicates and incompletes.
Although this is a rich heritage of human intellectual activities, it covers a very small portion of possible inorganic crystals.
Considering 82 non-radioactive chemical elements, the number of simple chemical compositions up to ternary compounds A$_a$B$_b$C$_c$ with integers satisfying $\max(a, b, c)\leq 15$ is approximately $10^8$, but increases to approximately $10^{10}$ for quaternary compounds A$_a$B$_b$C$_c$D$_d$.
Although many of these chemical compositions do not form stable crystals, the huge difference between the number of compounds in ICSD and the possible number of compounds implies that many unknown compounds remain.
Conventional experiments alone cannot fill this gap.
Often, first principles calculations are used as an alternative approach.
However, systematic first principles calculations without a priori knowledge of the crystal structures are very expensive.

Machine learning is a different approach to consider all chemical combinations.
A powerful machine-learning strategy is mandatory to discover new inorganic compounds efficiently.
Herein we adopt a recommender system approach to estimate the relevance of the chemical compositions where stable crystals can be formed [i.e., chemically relevant compositions (CRCs)].
The compositional similarity is defined using the procedure shown in Sec. \ref{nanoinfo:SecCompDes}.
A composition is described by a set of 165 descriptors composed of the means, SDs, and covariances of the established elemental representations.
The probability for CRCs is subsequently estimated on the basis of a machine-learning two-class classification using the compositional similarity.
This approach significantly accelerates the discovery of currently unknown CRCs that are not present in the training database.

\bibliography{nanoinfo}

\end{document}